\documentclass[acmsmall,nonacm]{acmart}
\usepackage{xcolor}
\usepackage{enumitem}
\usepackage{amsmath}
\usepackage{eucal}
\usepackage{fancybox}
\usepackage{xspace}
\usepackage{xcolor}
\usepackage[linesnumbered,ruled,lined]{algorithm2e}
\usepackage{url}

\SetCommentSty{mycommfont}

\newcommand{\so}[0]{Stack Overflow\xspace}
\newcommand{\blue}[1]{\textcolor{blue}{#1}}
\newcommand{\red}[1]{\textcolor{red}{#1}}

\usepackage{xcolor}
\usepackage{enumitem}
\usepackage{amsmath}
\usepackage{eucal}
\usepackage{graphicx}
\usepackage{booktabs}
\usepackage{xcolor}
\usepackage{url}
\usepackage[linesnumbered,ruled,lined]{algorithm2e}
\usepackage{balance}
\usepackage{multirow}
\usepackage{tcolorbox}

\newcommand{\se}[0]{software engineering}

\newcommand{\bench}[0]{\text{APISumBench}\xspace}

\newcommand{\pt}[0]{\text{PyTorch}\xspace}

\newcommand{\doc}[0]{\text{API documentation}\xspace}
\newcommand{\toolname}[0]{\textsc{APIDocBooster}\xspace}

\newcommand{\specone}[0]{\text{function}\xspace}
\newcommand{\spectwo}[0]{\text{parameters}\xspace}
\newcommand{\specthree}[0]{\text{notes}\xspace}

\newcommand{\compone}[0]{\text{CSSC}\xspace}
\newcommand{\comptwo}[0]{\text{UPSUM}\xspace}

\newcommand{\extalgo}[0]{\text{ExtUP}\xspace}

\newcommand{\llm}[0]{LLM\xspace}

\definecolor{babyblue}{rgb}{0.0, 0.5, 1.0}

\AtBeginDocument{%
  \providecommand\BibTeX{{%
    \normalfont B\kern-0.5em{\scshape i\kern-0.25em b}\kern-0.8em\TeX}}}

\begin{document}

\title{APIDocBooster: An Extract-Then-Abstract Framework Leveraging Large Language Models for Augmenting API Documentation}

\author{Chengran Yang}
\email{cryang@smu.edu.sg}
\author{Jiakun Liu}
\email{jkliu@smu.edu.sg}
\affiliation{%
  \institution{Singapore Management University}
  \country{Singapore}
}
\author{Bowen Xu}
\email{bxu22@ncsu.edu}
\affiliation{%
  \institution{North Carolina State University}
  \country{USA}
}
\author{Christoph Treude}
\email{christoph.treude@unimelb.edu.au}
\affiliation{%
  \institution{University of Melbourne}
  \country{Australia}
}
\author{Yunbo Lyu}
\email{yunbolyu@smu.edu.sg}
\author{Junda He}
\email{jundahe@smu.edu.sg}
\affiliation{%
  \institution{Singapore Management University}
  \country{Singapore}
}
\author{Ming Li}
\email{lim@lamda.nju.edu.cn}
\affiliation{%
  \institution{Nanjing University}
  \country{China}
}
\author{David Lo}
\email{davidlo@smu.edu.sg}
\affiliation{%
  \institution{Singapore Management University}
  \country{Singapore}
}





\vspace{-3mm}
\begin{abstract}

API documentation is often the most trusted resource for programming.
Many approaches have been proposed to augment API documentation by summarizing complementary information from external resources such as \so.
Existing extractive-based summarization approaches excel in producing faithful summaries that accurately represent the source content without input length restrictions. Nevertheless, they suffer from inherent readability limitations.
On the other hand, our empirical study on the abstractive-based summarization method, i.e., GPT-4, reveals that GPT-4 can generate coherent and concise summaries but presents limitations in terms of informativeness and faithfulness. 

We introduce \toolname, an extract-then-abstract framework that seamlessly fuses the advantages of both extractive (i.e., enabling faithful summaries without length limitation) and abstractive summarization (i.e., producing coherent and concise summaries).
\toolname
consists of two stages:
(1) \textbf{C}ontext-aware \textbf{S}entence \textbf{S}ection \textbf{C}lassification (\compone)
and (2) \textbf{UP}date \textbf{SUM}marization (UPSUM).
\compone classifies API-relevant information collected from multiple sources into API documentation sections.
\comptwo first generates extractive summaries distinct from original \doc and then generates abstractive summaries guided by extractive summaries through in-context learning.

To enable automatic evaluation of \toolname, we construct the first dataset for \doc augmentation. 
Our automatic evaluation results reveal that each stage in \toolname outperforms its baselines by a large margin.
Our human evaluation also demonstrates the superiority of \toolname over GPT-4 and shows that it improves the informativeness, relevance and faithfulness by 13.89\%, 15.15\%, and 30.56\%, respectively.

\end{abstract}



\maketitle








\section{Introduction}
\label{sec:intro}












The application programming interface (API) is one of the most vital components of modern application development.
Software developers typically rely on API reference documentation (API documentation in short) to learn APIs~\cite{treude2016augmenting,zhong2017empirical,uddin2019understanding}. 
API documentation is a set of documents indexed by the API name, where each document provides information about a specific API~\cite{maalej2013patterns}.
According to the 2023 \so Developer Survey~\cite{sosurvey},
technical documentation is the most trusted resource for programming.

However, creating and maintaining high-quality and readable API documentation still requires significant effort~\cite{aghajani2020software}.
Existing API documents are often incomplete and not always equally readable~\cite{uddin2015api,robillard2011field,aghajani2019software,treude2020beyond, maalej2013patterns}. 
A recent survey of professional developers shows that 60\% of the participants have suffered from inadequate API documentation in the last three months~\cite{khan2021automatic}. 
Motivated by this, researchers have proposed many solutions to augment API documentation~\cite{treude2016augmenting,wang2019extracting,uddin2017automatic, aghajani2019automated,zhang2019enriching}.


The state-of-the-art (SOTA) approaches SISE~\cite{treude2016augmenting} and DeepTip~\cite{wang2019extracting} formulate this task as \textit{update extractive summarization}.
Update summarization aims to generate complementary summaries to \doc, assuming readers are already familiar with the original API documentation~\cite{treude2016augmenting}.
Extractive summarization selects \textbf{insight sentences} for the target API from external sources to form extractive summaries.
Insight sentences provide API-relevant insights not covered in the API documentation~\cite{treude2016augmenting}. 
Nonetheless,  extractive summaries tend to show a notable level of redundancy and typically have limited readability~\cite{el2021automatic}. 
Notably, readability is one of the most important attributes of the desired software documentation~\cite{aghajani2020software, garousi2015usage, plosch2014value}, highlighting the need to produce concise and coherent summaries~\cite{aghajani2020software}.


Recently, research in utilizing large language models for abstractive summarization has garnered significant attention~\cite{zhang2023benchmarking, bhaskar2023prompted,zhang2023extractive}. 
Abstractive summarization generates new sentences, in contrast to extractive summarization which selects sentences from input.
The SOTA model GPT-4 demonstrates its potential in generating concise and coherent summaries~\cite{bhaskar2023prompted,zhang2023extractive}.
However, its effectiveness on API documentation augmentation remains unknown. 

In this work, we conduct the first empirical study on GPT-4 for \doc augmentation and compare it with existing approaches~\cite{treude2016augmenting,wang2019extracting} (detailed in Section \ref{sec:pilot}).
Our empirical study reveals that GPT-4 can generate coherent and concise summaries to augment \doc.
However, GPT-4 gives rise to several drawbacks concerning informativeness and faithfulness. 
First, the input length limitation of GPT-4 leads to information loss due to the truncation of the prompt.
We empirically discover that the token limit of GPT-4 allows for the inclusion of an average of only 6.35 \so threads relevant to an API to be embedded into the prompt.
Meanwhile, the cost of the GPT-4 API may be a burden for individual developers generating documentation (e.g., each GPT-4 API call costs around \$0.3 if the token number reaches the 8k limit).
Furthermore, GPT-4 occasionally generates non-faithful summaries, i.e., summaries that do not accurately reflect the original meaning of API-relevant resources~\cite{maynez2020faithfulness}. This phenomenon, known as hallucination~\cite{ji2023survey}, has hindered the trustworthy usage of GPT-4 in
industries like medicine~\cite{lee2023benefits} and law~\cite{cyphert2021human}.
We also observe that assessing the faithfulness of such hallucinations may require substantial time or, in some cases, might not even be feasible.
Since the accuracy of the information in software documentation is considered important for \se{} practitioners~\cite{aghajani2020software}, the existence of hallucinations also hinders the adoption of GPT-4 in this task.
Moreover, our pilot study reveals that the summaries generated by existing extractive summarization approaches~\cite{treude2016augmenting,wang2019extracting} are usually more faithful to external resources, which is attributed to direct sentence extraction without modification. Besides, extractive summaries demonstrate a considerable degree of informativeness and relevance, since there is no limitation on input length. 




The success of extractive summaries in terms of informativeness and faithfulness inspires us to use extractive summarization to guide abstractive summarization, addressing the drawbacks of the latter.
We leverage an extractive-then-abstractive pipeline including two phases: 1) Extract Phase: extract insight sentences from external resources to form extractive summaries, and 2) Abstract Phase: ask GPT-4 to generate abstractive summaries guided by extractive summaries.
The Extract Phase allows input without length limitations, covering more information from external resources and minimizing the costs for API calls.
The Abstract Phase ensures that abstractive summaries are aligned with extractive summaries, thereby enhancing faithfulness and facilitating data provenance.
We apply GPT-4 as the summarizer for the Abstract Phase.

We consider
SOTA extractive summarization approaches~\cite{wang2019extracting, treude2016augmenting} on the API documentation
augmentation task as candidates for implementing the Extract Phase.
However, we identify four drawbacks of existing approaches, which may hinder their performance in the Extract Phase. Therefore, we build upon existing approaches, extending them to overcome their limitations.
Notably, we are particularly concerned about the performance of the Extract Phase as the \textbf{extractive summaries directly determine the informativeness of the final \doc and their relevance to an API.}
Specifically, existing approaches
\textbf{1)} \textbf{Only take input from a single source.} They~\cite{treude2016augmenting, wang2019extracting} focus solely on \so, neglecting other sources such as tutorial videos and blogs that developers often refer to~\cite{sosurvey}. 
\textbf{2)} \textbf{Unaware of the API documentation structure.} 
Existing approaches perform binary classification to identify insight sentences, neglecting the standard structure of API documentation. The structure is important for software documentation~\cite{plosch2014value}, which commonly consists of multiple \textit{sections} (e.g., expected behavior, range of values, and cause of exceptions~\cite{javaspecification}).
\textbf{3) Neglect contextual dependencies.}
Existing approaches analyze each sentence individually and fail to capture semantic dependencies between sentences.
Consequently, insight sentences may be considered useless without necessary context, which leads to a 15\% output sentence confusion as reported by the existing approach SISE~\cite{treude2016augmenting}.
\textbf{4) Suffer from information redundancy.} \doc augmentation requires generated summaries to be distinct from the original \doc for readability~\cite{treude2016augmenting}. However, existing approaches neglect to reduce redundancy between generated summaries and \doc.

To address the drawbacks of existing approaches and fully leverage the capabilities of GPT-4, we propose \toolname, a novel framework with two stages:
\textbf{C}ontext-aware \textbf{S}entence \textbf{S}ection \textbf{C}lassification (\compone)
and \textbf{UP}date \textbf{SUM}marization (UPSUM).
\compone takes as input documents relevant to a specific API from \textbf{multiple sources}.
\compone identifies insight sentences and classifies them into suitable API documentation sections for \textbf{structure awareness} while considering the \textbf{contextual dependency} of sentences if necessary.
\comptwo applies an extract-then-abstract pipeline, which inputs insight sentences concerning a target API and outputs abstractive summaries to augment \doc.
In the Extract phase, 
we propose an \textbf{extractive update summarization algorithm} \extalgo that enables the generated summary to be semantically dissimilar from the API documentation.
In the Abstract phase, we leverage GPT-4 to generate summaries guided by the output of \extalgo through in-context learning. 

To enable automatic evaluation of this task, we construct the first dataset \bench through two-phase labeling. 
We identify 4,344 API-relevant sentences from multiple sources and linked them to appropriate \doc sections. 
Additionally, we produce 48 extractive summaries to augment three sections of \doc, corresponding to 16 APIs in Java and Python.

To assess \toolname, we conduct automatic evaluation using \bench on each stage of \toolname and human evaluation on end-to-end performance. 
Specifically, our automatic evaluation results reveal that \compone outperforms the best baseline by 18.18\%, 20.31\%, and 18.46\% in terms of precision, recall, and F1-scores.
Besides, \extalgo outperforms the best baseline by 14.55\%, 24.35\%, and 16.67\% in terms of ROUGE-1, ROUGE-2, and ROUGE-L.
Furthermore, our human evaluation results show that \toolname outperforms the GPT-4 baseline by 13.89\%, 15.15\%, and 30.56\% in terms of informativeness (i.e., quantity of insightful information), relevance (i.e., proportion of insightful information in generated summary), and faithfulness (i.e., to what extent the information in summaries aligns with corresponding information in external resources), as well as keeping on par performance on the readability (i.e., fluency and coherence) and the redundancy between the summaries and original API documentation.

The contributions of this paper are the following:
\begin{itemize}
    \item We conduct the first empirical study on GPT-4 performing abstractive summarization for API documentation augmentation. We identify three main drawbacks of GPT-4 and four drawbacks of existing extractive-based summarization approaches.
    \item We propose \toolname, a two-stage approach 
    with an extract-then-abstract framework to address these challenges. 
    \item We construct the first dataset \bench, which enables automatic evaluation of extractive summarization of \doc augmentation.
    \item We evaluate the performance of \toolname via both automatic evaluation and human evaluation. 
    Both evaluation results show that \toolname outperforms the best-performing baseline by a large margin. 
\end{itemize}

\section{PRELIMINARY}\label{sec:background}

\subsection{Motivation Example}
\label{sec:motivate}
In this section, we show an example of inadequate API documentation, how insightful information from external sources complements the original \doc, and the necessity for incorporating context information in Table~\ref{tab:adadelta}.

The first part of Table~\ref{tab:adadelta} shows an example of inadequate API documentation.
We present the functionality description of an API \textsc{Tanh}\footnote{\url{https://pytorch.org/docs/stable/generated/torch.nn.Tanh.html}} collected from PyTorch~\cite{pytorch}, a popular python framework for deep learning.
The description of the API \textsc{Tanh} in official \doc only contains a brief definition of the algorithm used in \textsc{Tanh} as well as the math formula. 
Such description discourages developers from better understanding the API due to its incompleteness, e.g., there is a lack of API usage, range of input and output, and information on quality attributes. Those aspects are considered important for \doc in the literature~\cite{maalej2013patterns}.

The second and third lines of Table~\ref{tab:adadelta} show that rich information relevant to the target API from external sources can complement the original \doc. 
We take the same API \textsc{Tanh} as an example.
The sentences extracted from the YouTube video caption present the usage scenarios and range of the output for \textsc{Tanh} API, while the quoted \so sentences highlight an essential property of \textsc{Tanh}. 
Both sentences can be considered a valuable complement to the API documentation. 
Furthermore, during our dataset collection process (as detailed in Section~\ref{sec:datasetstatistic}), we find that tutorial videos related to the target API often cover detailed operation steps of using the target API.
Meanwhile, the \so answers to ``how-to-do-it'' questions shed light on a target API's purpose and usage scenarios~\cite{treude2016augmenting}. 
Note that ``how-to-do-it'' questions are the most frequent question type on \so~\cite{de2014ranking}.
These observations inspire us to believe that the information contained in both YouTube videos and \so posts has great potential to augment API documentation and complement each other.

Furthermore, we show an example to illustrate the necessity of incorporating contextual information to ensure the comprehensiveness of API insights. Since existing approaches on \doc augmentation analyze each sentence individually by default, we argue this setting leads to a loss of API-relevant information and hinders readability.
We show an example in Table~\ref{tab:adadelta}.
The quoted sentences from Stack Overflow posts are semantically dependent on each other.
The first sentence illustrates that a function with a particular property can overcome the vanishing gradient problem. 
The second sentence explains that \textsc{Tanh} has such a property. 
Presenting either of these two sentences individually proves insufficient in conveying comprehensive insights concerning the target API.
Specifically, failing to consider the second sentence as the context of the first one leaves the connection between the first sentence and the target API unclear and obscure.
Meanwhile, presenting the second sentence in isolation fails to convey the detailed properties of the target API. This motivates us to incorporate contextual information when performing \doc augmentation.

\vspace{-2mm}

\begin{table}[t]
    \centering
    \caption{An Example of Inadequate API Documentation}
    \vspace{-4mm}
    \begin{tabular}{|p{8cm}|}\hline
        \textbf{TANH}  CLASS torch.nn.Tanh
        \\\hline
        \textbf{Functionality Description in Official Documentation }: \\
        Applies the Hyperbolic Tangent (Tanh) function element-wise.\\
        Tanh is defined as:\\
        \multicolumn{1}{|c|}{\(\displaystyle Tanh(x)=\frac{exp(x)-exp(-x)}{exp(x)+exp(-x)} \)}\\
        \hline
        \textbf{Related YouTube Video}: \\
        \small{\url{https://www.youtube.com/watch?v=lp4skE5U8Cs}}\\
        \dots 
        This Tanh is a hyperbolic tangent itself, but in the world of Neural Network, Tanh converts the input value into a non-linear one and keeps it in the range of -1\textasciitilde1.\\
        It is used to serve as an activation function.
        \dots \\
        \hline
        \textbf{Related \so post}: \\
        \small{\url{https://stackoverflow.com/a/40775144/10143020}}\\
        \dots 
        On the other hand, to overcome the vanishing gradient problem, we need a function whose second derivative can sustain for a long range before going to zero.\\
        Tanh is a good function with the above property.
        \dots \\
        \hline
    \end{tabular}
    \label{tab:adadelta}
    \vspace{-1mm}
\end{table}

\subsection{Notions and Task Formulation}
\label{sec:task}
This work aims to augment API documentation by generating complementary summaries for each documentation section from multiple online sources. 
As a initial attempt, we consider two sources, i.e., \so and YouTube, as input sources. The rationale is that existing work~\cite{sulistya2020sieve} reveals the potential of combining both platforms for software-related knowledge extraction.

Structuring is one of the most important quality attributes for effective software documentation~\cite{plosch2014value}. API documentation is typically divided into multiple smaller and more manageable sections. 
We consider three sections (i.e., Functionality, Parameter, and Notes) that appear in the API documentation of popular programming languages (e.g., Python~\cite{python} and Java~\cite{java}) as well as well-known third-party libraries (e.g., PyTorch~\cite{pytorch} and Android~\cite{android}). 
Note that some of the existing approaches~\cite{maalej2013patterns} categorize the knowledge required for \doc at a finer granularity. 
The reason for this coarse categorization is to better fit the structure of real-world \doc.
We leave the fine-grained \doc augmentation for future work.
We follow the definition of the API documentation section from Java API documentation's writing requirement~\cite{javaspecification}:

\begin{itemize}
    \item \textbf{Functionality}: 
    Information about the execution, expected behavior, state information and transitions, defined algorithms, and cause of exceptions of the API.

    \item \textbf{Parameters}: 
    Information about the range of valid argument values, return values, and behaviors of passing an invalid argument.
    \item \textbf{Notes}: 
    Information about OS/hardware dependencies, allowed implementation variances, security constraints, and references to any external specifications of the API, also other insightful tips (e.g., API performance, legal concerns) that are helpful for understanding an API~\cite{treude2016augmenting,wang2019extracting}.
\end{itemize}


\noindent Formally, the input for the \doc augmentation task includes a target API $api$, API documentation of the above three sections $doc$, and a set of sentences extracted from external sources $\mathit{s} = \{s_{1}, s_{2}, \dots, s_{n}\}$.
We aim to generate extractive summaries (i.e., an optimal subset ${sum}=Extract(s|api, doc), {sum} \subseteq \mathit{s}$), and then abstractive summaries (i.e.,  $sum^{'} = Abstract(sum|api, doc)$) that satisfy the requirements of readability, relevance of content, and structure of \doc. These attributes are the most critical to the quality of software documentation~\cite{plosch2014value}.

\subsection{Empirical Study on GPT-4}
\label{sec:pilot}
We carry out an empirical study to understand the capacity and limitations of using GPT-4 to augment API documentation.

\subsubsection{Statistics on External Resources.}
\label{sec:statistic}
We begin with exploring how much API-relevant information exists in external sources. 
Following existing approaches~\cite{chengran2022answer, uddin2017automatic}, we make an initial attempt on popular Python and Java libraries, i.e., \pt and Android. 
We randomly sample 100 APIs from each library, respectively. 
As discussed in Section~\ref{sec:task}, we consider \so and YouTube as preliminary external sources.

\so returns 258.70 posts on average for each API.
Corresponding \so threads account for 4,597.68 characters on average, which results in an average of approximately 1,150 tokens when tokenized by GPT-4~\cite{tokenizer}. The GPT-4 API imposes a maximum context limit of 8,192 tokens, inclusive of both prompts and the subsequent response. 
We reserve 500 tokens for API responses.
The summarization instruction and original API documentation together count 394 tokens on average.
Consequently, even if we consider Stack Overflow as the sole external source, only an average of 6.35 threads ($(8192-500-392)/1150$) can be included within the prompt for standard GPT-4, 27.05 threads for  GPT4-32k (i.e., GPT-4 variant with four times the context range and twice the API call cost).
Similarly, YouTube returns 80.98 videos on average.
The average length of the YouTube video captions is 41,816.49 characters, approximately 10,454 tokens. 
Even the average number of tokens for a single video exceeds the upper token limit of GPT-4's context.


\subsubsection{How does GPT-4 augment \doc?}
\label{sec:howgpt}
We instruct GPT-4 to automatically augment API documentation.
We set the parameter \textit{temperature} of GPT-4 as 0 to ensure GPT-4 produces a deterministic response. 
We consider two settings:

\noindent\textbf{Zero-shot Setting ($GPT_{zero}$).}
We ask GPT-4 to generate augmented \doc without external resources. 
We craft prompts that include original API documentation, the definition for each section, and an instruction for generation. We follow an existing GPT-based approach on opinion summarization~\cite{bhaskar2023prompted} to construct the instruction and prompts.\footnote{We present our prompts in the replication package due to the page limit.}

\noindent\textbf{Summarization Setting ($GPT_{sum}$).}
We ask GPT-4 to summarize insightful information for each section from given external resources. The generated summaries are required to be different from the original \doc. 
Specifically, we collect \so posts and captions of YouTube videos relevant to each API by calling the Stack Overflow search API and YouTube API (detailed in Section~\ref{sec:preprocessing}). The prompt used in this setting also follows existing approaches~\cite{bhaskar2023prompted, zhang2023extractive} and consists of original API documentation, the definition for each section, an instruction for summarization, and API-relevant resources.




\vspace{-1mm}
\begin{table}[htbp]
    \centering
    \caption{User Study for GPT-4 (5-point Likert scale)}
    \vspace{-4mm}
    \resizebox{\columnwidth}{!}{
    \begin{tabular}{cccccc}
    \toprule
    \textbf{System} & \textbf{Informativeness.}  & \textbf{Relevance}  & \textbf{Readability} & \textbf{Non-Redundancy} & \textbf{Faithfulness} \\
    \midrule
    $GPT_{zero}$ & 2.6 & 2.2 & 4.3 & 4.5 & - \\
    $GPT_{sum}$ & \textbf{3.6} & \textbf{3.3} & \textbf{4.4} & \textbf{4.5} & 3.6 \\
    SISE & 3.0 & \textbf{3.3} & 2.3 & 2.6 & \textbf{5.0} \\
    \bottomrule
    \end{tabular}
    }\\
    \label{table: userstudy1}

\end{table}

\vspace{-5mm}
\subsubsection{User Study.\\} 
\noindent \textbf{User Study Setting.}
We leverage $GPT_{zero}$, $GPT_{sum}$, and SISE (i.e., SOTA extractive-summarization-based approach~\cite{treude2016augmenting}), to augment \doc.
We randomly sample 10 APIs from \pt and Android libraries respectively.
As the length of a single video caption exceeds the token limit of GPT-4, we only consider \so as external sources for $GPT_{zero}$ and $GPT_{sum}$. We add each post in the prompt in the same order that they are returned by \so, until reaching the context limitation of GPT-4. 
We ask GPT-4 to generate summaries for all \doc sections. 

Four user study participants, with at least two years of Python or Java programming experience, read all augmented API documentation pages and rate each section on a 5-point Likert scale in terms of \textbf{informativeness}, \textbf{relevance}, \textbf{readability}, \textbf{non-redundancy}, and \textbf{faithfulness}. 
Informativeness represents the quantity of insightful information; relevance measures the proportion of insightful information; readability assesses the fluency and coherence of the summaries; non-redundancy evaluates the information redundancy between the summaries and original \doc;
and faithfulness is a measure of the degree to which information in summaries aligns with corresponding information in external resources. 
Scores 1 to 5 represent a range from extremely low to extremely high, respectively.
All of the evaluation metrics are widely used in the \se{} domain ~\cite{treude2016augmenting,chengran2022answer} and in NLP domain summarization approaches~\cite{huang2020have,bhaskar2023prompted}.
Participants are also required to highlight sentences that they think should not be in \doc given the above criteria.
Note that, summaries with hallucinated information are given a lower faithfulness score.
However, it is impossible to assess the faithfulness of $GPT_{zero}$ as GPT-4 generates the summaries without external resources as reference.

\noindent \textbf{User Study Results.}
As shown in Table~\ref{table: userstudy1}, $GPT_{sum}$ performs best in terms of informativeness, readability, and redundancy. 
SISE performs best in terms of faithfulness and shows on-par relevance compared with $GPT_{sum}$. 
The superior performance of $GPT_{sum}$ compared with $GPT_{zero}$ in terms of informativeness (p<0.01) indicates the necessity of leveraging external resources. The lower performance of SISE compared with the GPT-based approach in terms of redundancy and readability (p<0.01) demonstrates the potential of GPT-4 to generate fluent and coherent summaries.

\noindent\textbf{Qualitative Analysis.}
Two authors of this paper also perform qualitative analysis on the augmented \doc generated by $GPT_{zero}$ and $GPT_{sum}$. Specifically, they go through the sentences highlighted by participants of the user study and discuss the rationale behind the unsuitability of these sentences within the \doc context.
Firstly, we observe that $GPT_{sum}$ occasionally produces less informative sentences.
For instance,
GPT-4 generated the following sentence: \textit{``Implementation may vary across different OS/hardware.''} for the OS dependency of API LeakyReLU in PyTorch~\cite{pytorch}. 
Upon seeking further explanation from GPT, it clarifies that the implementation differences arise from hardware-specific optimizations of PyTorch on various platforms. 
This information applies to most APIs and lacks specificity to the target API, resulting in a sentence with minimal insights to \doc.
We attribute this to the limited availability of high-quality API documentation examples provided to GPT, which hinders GPT's ability to learn from document examples and identify the most insightful information specific to a particular API.
However, the input length limitation of GPT-4 makes it challenging to feed more \doc examples.

We also observe hallucinations, that is information not faithful to given resources, in the generated summaries. 
Specifically, $GPT_{zero}$ consistently generates incorrect parameter lists, including parameters from APIs with similar names but in different libraries, when original API documentation is not provided. Even when the API documentation is embedded in the prompt, there are instances where the $GPT_{zero}$ still confuses the numerical range of the parameters. We have also noticed a reduction in hallucinations with the use of the $GPT_{sum}$, although there are occasional occurrences of $GPT_{sum}$ generating information that is not accurate compared with given external resources. On the contrary, SISE does not generate hallucinations since it only extracts sentences from given external resources.
Based on the above findings, we argue that providing more external resources can help alleviate hallucinations of GPT-4, enabling GPT-4 to summarize from given information rather than generate information itself.

\vspace{-1mm}
\subsection{Usage Scenario}





We propose \toolname as an automated aggregator of API-related information and writer to update API documentation.




More specifically, with the help of \toolname, API documentation's maintenance and update processes become more streamlined and efficient.
\toolname simplifies the tedious task of documentation maintenance~\cite{aghajani2020software} into an automatic pipeline: 1) automatically collecting API-related information from external sources; 2) automatically generating summaries for each section of the API documentation; and 3) integrating the generated summaries into the original API documentation.

Notably, insights provided by \toolname highlight the characteristics of the target API that developers are concerned about in practice. The focus of the developer community is commonly mirrored in the frequency of corresponding information mentioned in external resources such as Stack Overflow~\cite{uddin2017automatic}. As the extractive summarization algorithm prioritizes information based on its mention frequency, \toolname enables API-related information with substantial community attention more likely to be involved in the summary. 
Consequently, \toolname enables \doc maintainers to focus on updating information that more developers are interested in, which in turn, accommodates the informational needs of a more diverse readership.

It is essential to acknowledge that \toolname's effectiveness relies on the richness of API-related discussions within the development community. APIs with limited discussions might present constraints in \toolname's performance. However, this limitation is deemed acceptable given \toolname's objective: to enhance \doc that are highly demanded for augmentation. 
We argue that the existence of extensive community discussions concerning specific APIs implies deficiencies within the original API documentation.
The rationale is that technical documentation is the primary way for developers to acquire coding knowledge as indicated by Stack Overflow's survey~\cite{sosurvey}, thereby a sufficiently comprehensive API document would ensure a profound comprehension of API usage for the majority of developers, obviating the necessity for external inquiries. 
Consequently, an increased frequency of community discourse concerning a specific API indicates a higher probability of inadequacies in its documentation. This, in turn, underscores the imperative to augment the quality of said documentation.
\section{Approach}
\begin{figure*}
    \centering
    \includegraphics[width=\columnwidth]{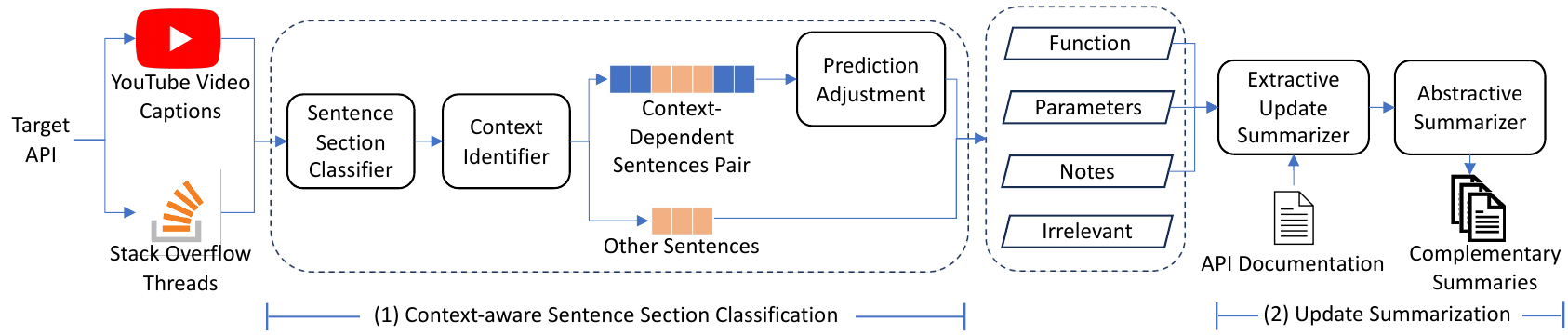}

    \caption{Overview of \toolname. (1) we train a context identifier and a sentence section classifier to identify the context of each sentence and divide each sentence into either one of the three sections or API-irrelevant, respectively. (2) we generate extractive summaries for each section by using our proposed update summarization algorithm. We then generate abstractive summaries guided by extractive summaries.}
    \label{fig:overview}
\end{figure*}

\begin{figure*}
    \centering
    \includegraphics[width=\columnwidth]{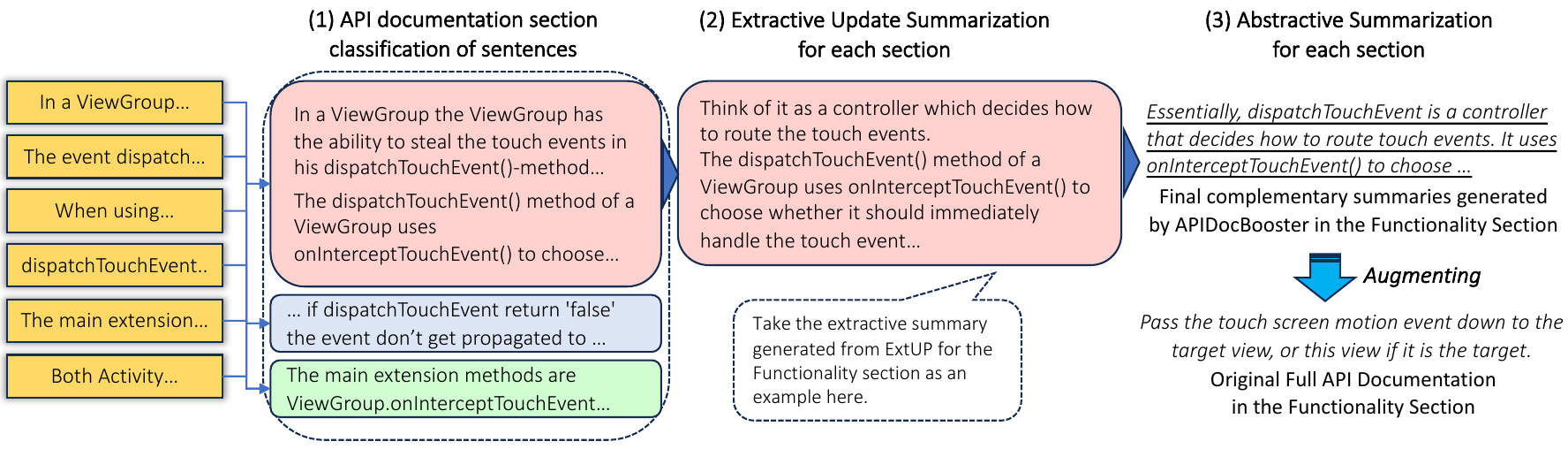}

    \caption{Example of \toolname's summarization process. (1) Given sentences extracted from external sources, we train a sentence section classifier to classify sentences into either one of three \doc sections (i.e., \red{functionality}, \blue{parameter}, \textcolor{green}{notes}) or API-irrelevant one. Then we leverage our extractive update summarization algorithm to generate extractive summaries for each section. Finally, we ask a large language model to generate abstractive summaries for each section guided by extractive summaries.}    
    \label{fig:sum_example}
\end{figure*}

\subsection{Overview}

The framework of our approach, namely \toolname, is shown in Figure~\ref{fig:overview}.
\toolname takes as input \so posts and YouTube videos for a given API.
\toolname outputs a set of abstractive summaries, each of them corresponding to a specific API documentation section.
\toolname contains two core components: 
1) \textbf{C}ontext-aware \textbf{S}entence \textbf{S}ection \textbf{C}lassification (\compone)
and 2) \textbf{UP}date \textbf{SUM}marization (UPSUM).
\compone takes as inputs \so posts and YouTube video captions and outputs sets of sentences predicted to be insightful to each section while taking sentence context into account.
Given a list of insightful sentences for an API documentation section, 
\comptwo follows extract-then-abstract pipeline and generates abstractive summaries.
In the extract phase, we propose an \textbf{Ext}ractive \textbf{up}date summarization algorithm (\extalgo) to generate extractive summaries. Then \comptwo leverage GPT-4 to generate abstractive summaries guided by the output of \extalgo.
We also show an example of the processing of data flow in Figure~\ref{fig:sum_example}.

\vspace{-2mm}
\subsection{Pre-processing}
\label{sec:preprocessing}
In the pre-processing step, we \so posts and YouTube video captions relevant to the given API. 

\noindent \textbf{Searching for API-relevant Resources.}
To extract Stack Overflow posts relevant to specific API, we call the Stack Exchange API~\cite{stackexchange}. 
We search with the query $library\_name + API\_name$ and consider returned answer posts with at least one up vote as posts relevant to the target API. 
We also filter out \so answer without textual content. 
To obtain the YouTube videos relevant to the target API, we use the official YouTube API~\cite{youtubeapi} to search for videos with the same query as for \so Exchange API search.

\noindent \textbf{Sentence Extraction.}
We apply NLTK's sentence tokenizer~\cite{senttokenizer} to split the textual content from \so posts and captions of YouTube videos into sentences.\footnote{YouTube caption files contain the text of what is said in the video. It also contains timestamps for when each line of text should be displayed. } 
Furthermore, since video caption has much more API-irrelevent content compared to \so answers, we denoise the caption data. 
As described in Section~\ref{sec:statistic}, YouTube video captions have 9.10 times more characters than \so posts on average.
However, the majority of the sentences are extraneous to the specific APIs. 
In our manually labeled dataset (detailed in Section~\ref{sec:benchmark}), the average number of insight sentences for the target API, among 553 sentences per video, is 11.9.
To reduce noise in input, inspired by existing multi-doc summarization for question and answering~\cite{xu2020coarse}, we leverage the BM25~\cite{robertson2009probabilistic} algorithm to coarsely filter out video captions sentences that are irrelevant to API.
BM25 is a ranking function that is used to estimate the relevance between a given query and a document. Formally, given a query $Q$, the relevance score is given by:
\begin{equation}
    \small
        Score(Q,d)=\sum_{i}^{n}IDF_{q_{i}}\frac{h(q_{i},d)*(k_{1}+1)}{h(q_{i},d)+k_{1}*(1-b+b*\frac{\left | D \right |}{avgdl})}
    \end{equation}

\noindent where $q_{i}$ is the $i^{th}$ word in the query $Q$, $h(q_{i},d)$ indicates the frequency of $q_{i}$ occurring in the document $d$, $avgdl$ is the \textbf{a}\textbf{v}era\textbf{g}e \textbf{d}ocument \textbf{l}ength in the text collection, $k_{1}$ and b are adjustable parameters (by default $k_{1}=1.2$ and $b=0.75$, following Robertson et al.~\cite{robertson2009probabilistic}), and $\left | D \right |$ is the length of the document in terms of words. The $IDF(q_{i})$ is the IDF value of $q_{i}$. 
We observe that sentences, along with their context, which contain name entities within \doc (e.g., API name and API parameters), tend to be insightful to target API.
Thus, we set each caption sentence as $d$ and manually construct the query for each API as \textit{``API name, hyper-parameter names, input and output''}. 
Following existing work~\cite{xu2020coarse}, we consider sentences with positive BM25 scores as well as their context (i.e., preceding and following sentence) as input for \compone. We consider preceding and following sentences as context since adjacent sentences are more likely to have semantic dependencies.

\subsection{Context-aware Section Classification}
\compone trains a BERT-based classifier to classify sentences from relevant \so answers and YouTube video captions as either one of the three section types (i.e., \specone, \spectwo, \specthree) or API-irrelevant.
In addition, we implement a context awareness mechanism to improve the classification performance of context-dependent sentences.

\setlength{\textfloatsep}{3pt}
\begin{algorithm}[htbp]
\small
    \caption{Context-Awareness Mechanism}\label{alg:summarygeneration}
    \SetKwComment{Comment}{/* }{ */}
    \SetKwInput{KwInput}{Input}                
    \SetKwInput{KwOutput}{Output}
    \KwInput{
    $api$: target API;    
    S: List of the API-relevant sentences\\}

    \KwOutput{$\mathit{S}\_\mathit{label}: section~label~of~each~sentence$}
    \textbf{Initialization}:
    $\mathit{context}\_\mathit{flag}$ = {[]};
    $l^{update}$ = ""



\For{$s_{i-1}$,$s_{i}$,$s_{i+1}$  \textbf{in} S}{

\eIf{$\mathit{context}\_\mathit{identifer}$($s_{i-1}$,$s_{i}$)}
{$\mathit{context}\_\mathit{flag}\_\mathit{preceding}${[i]} = True\\
$\mathit{context}\_\mathit{score}\_\mathit{preceding}$[i]=$\mathit{context}\_\mathit{identifer}$($s_{i-1}$,$s_{i}$)
}
{$\mathit{context}\_\mathit{flag}${[i]} = False}
\eIf{$\mathit{context}\_\mathit{identifer}$($s_{i}$,$s_{i+1}$)}
{$\mathit{context}\_\mathit{flag}\_\mathit{following}${[i]} = True\\
$\mathit{context}\_\mathit{score}\_\mathit{following}$[i]=$\mathit{context}\_\mathit{identifer}$($s_{i}$,$s_{i+1}$)
}
{$\mathit{context}\_\mathit{flag}${[i]} = False}
}

    \For{$s_i$ \textbf{in} S}{
        $s_{context}$ = s[i]\\
        \If{$\mathit{context}\_\mathit{flag}\_\mathit{preceding}${[i]}== True}{$s_{context}$=$s_{context}$+$s_{i-1}$}
        \If{$\mathit{context}\_\mathit{flag}\_\mathit{following}${[i]}== True}{$s_{context}$=$s_{context}$+$s_{i+1}$}
        \If{$\mathit{context}\_\mathit{flag}\_\mathit{preceding}${[i]}== True \textbf{or} $\mathit{context}\_\mathit{flag}\_\mathit{following}${[i]}== True}
        {
        $l^{origin}$ = section\_classify($s_i$,$api$)\\
        $l^{context}$ = section\_classify($[s_{i}, s_{context}]$,$api$)\\
        $l^{update}$ = update\_section\_prediction( $l^{origin}$,$l^{context}$)
        }
    }

    sent\_labels = argmax($l^{update}$)
        
    
\end{algorithm}

\subsubsection{Sentence Section Classifier}
\label{sec:apiblockclassifier}
We fine-tune BERT to create a four-class classifier to divide each input sentence into either one of the three API documentation sections or irrelevant.
We leverage BERT, a pre-trained language model that achieves promising performance on relevant software engineering tasks~\cite{chengran2022answer,yang2022aspect}, to fine-tune a sentence section classifier. 
Previous approaches~\cite{treude2016augmenting,wang2019extracting} treat the task as a binary classification that identifies insight sentences to the target API. 
Unlike previous approaches, we aim to further classify insight sentences into a finer granularity, i.e., sections of the API document. Thus we perform multi-class classification.
By doing so, each sentence is mapped to a certain document section, enabling augmented API documentation at a more fine-grained level. 

In our task, an input instance is in the form of a pair of API and corresponding sentences, i.e., $\langle API, S\rangle$; 
thus, we feed the pairs of sentences and APIs to the BERT model for fine-tuning. 
Specifically, for each pair $\langle S, API\rangle$, we concatenate the sentence $S$ and the target API $API$ into a sequence with padding if necessary, i.e., $\langle[CLS],S,[SEP],API,[SEP]\rangle$, to serve as input to the classifier. 
Both $[CLS]$ and $[SEP]$ tokens are used to form the BERT input~\cite{devlin2018bert}. 
The $[CLS]$ vector extracted from the classifier decoder is fed into the final classification layer to produce the prediction. 
Following standard practice, we use cross entropy as the loss function. 

\vspace{-3mm}
\subsubsection{Context-Awareness Mechanism}
\label{sec:context-awareness}
As we discussed in Section~\ref{sec:motivate}, we argue that the necessary context of a given sentence plays a key role in the understanding of this sentence and delivering complete information. 
Sentences requiring contextual information for the comprehension of API insights are denoted as context-dependent sentences.
However, not all sentences need context to be understandable, where unnecessary context is considered as noise. 
Thus, we propose a context-awareness mechanism to 1) determine context-dependent sentences, 2) concat context-dependent sentences with necessary context, and 3) modify the section classification results for context-dependent sentences based on their context.
We detail the algorithm of the context-awareness mechanism shown in Algorithm~\ref{alg:summarygeneration} as follows.

\noindent \textbf{Training Context Identifier}.
Firstly, we train a context identifier to identify context-dependent sentences.
To train such a context identifier, we fine-tune a binary BERT-based classifier to model the context dependence for a pair of consecutive sentences.
The classifier inputs a pair of consecutive sentences and outputs whether they are contextually dependent on each other.
To train such a model, we reuse the training data of
our dataset \bench (described in Section~\ref{sec:labelingprocess}).
In \bench, annotators label consecutive sentences with context-dependence as the same semantic unit.
We reuse the context information inferred by the semantic unit label to construct positive examples.
Precisely, we extract pairs of consecutive sentences in the same semantic unit as positive samples (i.e., label the pair with 1).
To construct the negative pairs (i.e., label the pair with 0), we sample pairs of sentences randomly selected from different Stack Overflow answers or YouTube videos.
We fine-tune the BERT-based classifier with the above dataset. 
We apply the same fine-tuning strategy as our sentence section classifier.
Following standard practice, we use cross entropy as the loss function. 

\noindent \textbf{Context Dependency Prediction (Line 2-15)}.
After we obtain our context identifier model, we predict if each sentence input in \compone is context-dependent. 
For a given sentence, we only consider its preceding and following sentence as the candidates of context-dependent.
The rationale is that semantic dependencies are more probable between adjacent sentences.
Next, a sentence is predicted to be context-dependent if at least one of its preceding and following sentences has contextual dependence with this sentence.
Besides, we pick the positive class' prediction probability $\textbf{c}$ as each pair's context-dependent score, indicating the extent to which the pair of sentences are semantically dependent on each other.

\noindent \textbf{Context-aware Prediction Adjustment (Line 16-29)}. 
Once a sentence $s_{i}$ is predicted as context-dependent, we update its prediction logit $l = \left[l_{1}, l_{2}, l_{3}, l_{4}\right]$, i.e., raw prediction score for API documentation section classification,  by considering the prediction that the input sentence is concatenating with context.
The value $l_{i}$ is a raw prediction score (i.e., the score before normalization) of each class. 
Firstly, we obtain the prediction logits $l^{origin}$ of $s_{i}$ by feeding $s_{i}$ and corresponding the API $api$ to the section classifier:

\vspace{-3mm}
\begin{equation}
\small
    l^{origin} = \mathit{section}\_\mathit{classify}(s_{i},api)
\end{equation}
\vspace{-3mm}

\noindent where $section\_classify$ denotes the section classifier.
Then, assuming $s_{i-1}$ and $s_{i+1}$ are contextual sentences of the current sentence $s_{i}$, we obtain the context-aware prediction logits of $s_{i}$ by concatenating $s_{i-1}$, $s_{i}$, and $s_{i+1}$ and feeding them to the section classifier:

\vspace{-3mm}
\begin{equation}
\small
    l^{context} = \mathit{section}\_\mathit{classify}([s_{i-1}, s_{i}, s_{i+1}],api)
\end{equation}
\vspace{-3mm}

Next, we update the prediction logits $l^{update}$. 
Specifically, in each dimension of $l^{update}$, the value $l_{j}^{update}$ is obtained by weighting both context-aware prediction logits $l_{j}^{context}$ and original prediction logits $l_{j}^{origin}$:

\vspace{-3mm}
\begin{equation}
\small
    l_{j}^{update} = \frac{\sum ((1-c) \times l_{j}^{origin} + c \times l_{j}^{context})}{num(context)}
\end{equation}
\vspace{-3mm}

\noindent where $num(context)$ denotes the number of contextual sentences of the current sentence. 
$\textbf{c}$ represents the extent of semantic dependency between the current sentence and its contextual sentences.
Finally, given update prediction logits $l^{update} = \left[l_{1}^{update}, l_{2}^{update}, l_{3}^{update}, l_{4}^{update}\right]$, we obtain the update predicted section $y'$ by applying the $argmax$ function:
\begin{equation}
\small
    y' = \mathit{argmax}(l^{update})
\end{equation}

\subsection{Update Summarization}
Given the output of \compone (i.e., a set of insightful sentences for each API documentation section) and a budget of summarization length (i.e., number of sentences) as input,
\comptwo outputs a set of summaries for augmenting API documentation sections.

Specifically, \comptwo employs an extract-then-abstract summarization pipeline, in which the abstractive summarization is guided by the output of extractive summarization. 
Extractive summarization involves the direct selection of sentences from the source text, ensuring consistency of information and establishing a clear lineage between the source and the generated summaries~\cite{el2021automatic}.
Conversely, abstractive summarization entails generating summary content from the model, resulting in a fluent summary but potentially containing words not present in the source and opinions contradicting the source (i.e., hallucination)~\cite{el2021automatic}.
The purpose of the extract-then-abstract pipeline is threefold: 1) to link the generated content in abstractive summaries with corresponding external sources, 2) to take advantage of both extractive (to enhance faithfulness and mitigate hallucination) and abstractive summarization (to enhance readability), and 3) to alleviate the length constraints inherent in abstractive summarization models.

Precisely, in the extractive summarization phase, \comptwo performs extractive update summarization. Update summarization is a double-objective summarization task~\cite{treude2016augmenting}, which aims to generate a summary by minimizing redundancy 1) across generated content and 2) between generated summaries and original API documentation.
We achieve this goal by utilizing our proposed extractive update summarization algorithm \extalgo for update summarization.
In the abstractive summarization phase, \llm performs in-context learning and generates abstractive summaries guided by the extractive summaries. 

\subsubsection{Extractive Update Summarization}
We propose an extractive update summarization approach, \extalgo.
Specifically, given a target API, \extalgo inputs a set of sentences that are insightful to the specific \doc section. \extalgo outputs an optimal subset of sentences as the summary that represents the main ideas of the input while minimizing the information redundancy between the summary and original API documentation.
We modify a simple yet effective summarization approach, TextRank~\cite{mihalcea2004textrank}, for update summarization.

TextRank is an unsupervised extractive summarization approach based on the PageRank algorithm. 
PageRank is interpreted as the stationary distribution of a random walk on a graph that restarts from a uniform location in the node sets at each time.
The restart probability is $1-\phi$ while the probability of moving to a neighbor node is $\phi$.
In TextRank, all input sentences are constructed as a sentence graph with undirected edges.
Each node represents a sentence, and the initial weight of an edge is based on the token similarity between two nodes.
TextRank recursively computes the similarity of each sentence to all candidates as follows:

\begin{equation}\small
    C(S_{i})=(1-\phi) +\phi\times\sum_{V_{j}\in In(V_{i})}^{}\frac{1}{Out(S_{i})}C(S_{j})
\end{equation}

To incorporate the update mechanism in TextRank, we add a penalty factor $p_{doc}$ to deprioritize sentences that are semantically similar to original API documentation when performing summarization. This penalty factor is inspired by Biased-TextRank~\cite{kazemi2020biased}. Biased-TextRank is designed for query-focused summarization, which is biased toward query-oriented sentences. 
Specifically, our penalty factor changes the way random restart probabilities $(1-\phi)$ are assigned in the sentence graph.
In TextRank, all nodes have an equal restart probability of $1-\phi$.
However, in our scenario, sentence nodes that are semantically similar to the \doc are expected to have a lower restart probability.
Nodes with lower random restart likelihood are less likely to be reached when randomly jumping, and thus achieve lower Summarization scores.  
The penalty factor is defined as follows:
\begin{equation}\small
    p_{doc} = 1-cosine\_similarity(node, APIdoc)
\end{equation}
\noindent where $node$ and $APIdoc$ denote the sentence representation in node graph and original \doc, respectively. 
We apply a software-engineering domain text representation model~\cite{chengran2022answer} to obtain their representations. The formulation of calculating sentences' ranking is defined below, which is followed by a normalization:
\begin{equation}\small
    C(S_{i})=p_{doc}\times(1-\phi) +\phi\times\sum_{V_{j}\in In(V_{i})}^{}\frac{1}{Out(S_{i})}C(S_{j})
\end{equation}
\noindent where $In(S_{i})$ represents all nodes that point to $S_{i}$ and $Out(S_{i})$ represents all nodes that $S_{i}$ points to. 
$C(S_{i})$ denotes the TextRank summarization score of a node sentence.
$\phi$ is the damping factor, defaulting as 0.85~\cite{mihalcea2004textrank}.

We also replace the way of assigning the edge weight in the sentence graph from calculating token overlaps to calculating the cosine similarity between two sentences' representations.
We use the same text representation model~\cite{chengran2022answer} as calculating the penalty factor $p_{doc}$. 
Finally, given the budget of summary length, i.e., $k$ sentences, we select the top-k sentences with the highest scores as the extractive summary.

\subsubsection{Abstractive Summarization}
We leverage GPT-4 to generate abstractive summaries guided by the extractive summaries.
Specifically, we ask GPT to select insightful information from the extractive summaries (i.e., $S_{e}$) concerning the specification of API sections, paraphrase and synthesize the insightful information, and form abstractive summaries (i.e., $S_{a}$).
We follow existing GPT-based approaches for opinion and news summarization to construct the prompts~\cite{bhaskar2023prompted, zhang2023extractive}, including extractive summaries $S_{e}$, the summarization instruction $i$, original API documentation $d$, and context pairs $\{(S_{a1}, S_{e1})...(S_{an}, S_{en})\}$.\footnote{Detailed prompts are provided in our repository due to the page limit.} 
We link the sentences in extractive summaries and abstractive summaries by greedily calculating the cosine similarity between sentence representations~\cite{yang2022answer}.
Formally, the summarization process is represented as: 

\begin{equation}\small
    \hat{S_{a}} =\underset{S_{a}}{argmax}P_{GPT}(S_{a}|S_{e},d,i,\{(S_{a1}, S_{e1})...(S_{an}, S_{en})\})
\end{equation}
 

As we observe that GPT-4 tends to generate less informative content in Section~\ref{sec:pilot}, we perform in-context learning to enhance the informativeness of generated summaries. In-context learning can make \llm rapidly adapt to or recognize the desired task by providing example input-output pairs $\{(S_{a1}, S_{e1})...(S_{an}, S_{en})\}$ in the prompt for few-shot learning~\cite{brown2020language}. We provide pairs of ground-truth extractive summaries and informative abstractive summaries, aiming to make GPT-4 more adept at producing informative summaries. 
Specifically, we sample human-crafted extractive summaries from our dataset \bench{} (detailed in Section~\ref{sec:benchmark}) and ask GPT-4 to generate augmented \doc following the same prompt. 
We then manually revise the augmented \doc to remove less informative content.

\section{Experiment Setting}

\subsection{\bench}
\label{sec:benchmark}
We craft \bench{}, the first dataset which enables automatic evaluation of \compone and \extalgo.
In \bench, we perform two-stage labeling and acquire two phases of data.
In the first phase, we invite annotators to label sentences from \so posts and YouTube videos relevant to the target API with API documentation sections.
The first phase consists of 4,344 sentences, which scale is on par with previous API review summarization work~\cite{uddin2017automatic}.
The second phase comprises 48 extractive summaries, generated from the data gathered in the first phase, and corresponds to three API documentation sections from 16 different APIs.
The first phase data enables automatic evaluation on \compone, i.e., sentence classification for \doc section.
The second phase data enables automatic evaluation on \extalgo, i.e., extractive summarization for API documentation augmentation. 

\subsubsection{Data Collection.}
We collect candidate APIs as well as relevant \so posts and YouTube videos.
Following our empirical study in Section~\ref{sec:pilot}, we randomly sample 10 \pt APIs and 10 Android APIs as candidates.
To reduce the labor cost, following existing approaches~\cite{chengran2022answer}, We collect the top 10 \so posts returned by \so search engine for each API.
We notice that numerous Stack Overflow posts, which are identified as relevant to the target API by the Stack Overflow search engine, solely contain API in the code snippets. Those posts lack insight sentences towards the target API.
Thus, we filter out posts containing target API only in code snippets. 
For YouTube video captions, we observe that automatically generated captions have no punctuation, thus causing those captions to be unable to be naturally split into sentences.
To improve the data quality in \bench, we only automatically extract videos with manually crafted captions. 
We apply the youtube-transcript-api tool~\cite{transcript} to automatically extract videos with manual captions. 
In this step, 189 \so posts and 84 YouTube videos are collected, i.e., 11.72 \so posts and 4.2 YouTube videos for each API on average. 
We then use NLTK~\cite{NLTK} to break the collected \so posts and YouTube videos into sentences. For YouTube video captions, we follow our pre-processing step in Section~\ref{sec:preprocessing} to reduce the noise. 
Finally, we obtain 5,414 sentences related to 20 APIs. 

\subsubsection{Labeling Process.}
\label{sec:labelingprocess}
Following prior works~\cite{chengran2022answer, angelidis2018summarizing},
we perform two-phase labeling.
Four participants are involved in the labeling process.
None of them participated in the user study detailed in Section~\ref{sec:pilot}.
All annotators have at least two years of programming experience in Java and Python. 

\noindent\textbf{Phase 1: API Documentation Section Classification.}
We ask annotators to label both API documentation sections and the context dependence of each sentence \cite{treude2016augmenting}. 
Specifically, annotators cluster consecutive sentences that are context-dependent as a cluster. 
Annotators consider a cluster as a single sentence in the following labeling. 
Then they label each sentence with \doc sections.
Sentences in the same cluster are labeled with the same section label. 
The whole labeling process consists of 3 steps:\\
\textbf{1) Iterative Guideline Refinement.} 
We iteratively refine the guideline for annotators before full labeling. 
Specifically, we ask annotators to label the data for one API following the guideline. We then calculate their labeling agreement in terms of Kappa value\footnote{We interpret kappa values 0 as poor agreement, 0.01–0.20 as slight agreement, 0.21–0.40 as fair, 0.41– 0.60 as moderate, 0.61–0.80 as substantial, and 0.81–1.00 as almost perfect agreement~\cite{donker1993interpretation}.}. 
Next, we refine the guideline based on the annotators' feedback.
We repeat the process until the Kappa value reaches a substantial level of agreement (i.e., 0.62), which agreement surpasses common summarization datasets in software engineering domain~\cite{chengran2022answer} and NLP domain~\cite{angelidis2021extractive}. Finally, the data for four APIs used in the refinement step are dropped due to the low agreement.\\
\noindent\textbf{2) Full Labeling.}
Annotators then follow the refined guideline to label the remaining 16 APIs in the full labeling process.
Each annotator independently labels sentences for 4 APIs.
In total, we label 4,344 sentences with API documentation sections in the first labeling stage.\\
\noindent\textbf{3) Validation Labeling.}
We perform validation labeling to assess the quality of the data.
We randomly sample 654 sentences from the full labeling data (i.e., 15\%) and the first author labels the sampled sentences independently.
The Cohen's Kappa score between the annotators and the first author achieves a substantial level of agreement (i.e., 0.61), which is higher than that of existing summarization datasets in the NLP domain~\cite{angelidis2021extractive} and Software Engineering domain~\cite{chengran2022answer}.

\noindent \textbf{Phase 2: Update Summarization.}
Following existing work~\cite{chengran2022answer}, in the second phase, 
given a set of sentences associated with each API documentation section, annotators select at most five sentences to form a summary for each section,
considering the insights not covered in the target API's \doc~\cite{treude2016augmenting} and the redundancy among all sentences.
Phase 2 employs the same guideline refinement step and full labeling step as phase 1.
Finally, 48 ($16 APIs \times 3 sections$) extractive summaries are produced. 

\subsubsection{Dataset Statistics}
\label{sec:datasetstatistic}
In \bench{}, the first phase data consists of 4,344 labeled sentences labeled with API documentation sections or irrelevant to the API. 
The average number of sentences input into the second stage for each section (i.e., function, parameter, and notes) is 18.76, 4.33, and 11.08, respectively.
The average number of sentences in the generated summaries of each section is 4.69, 2.77, and 4.31.
On average, 20.51\% of sentences in the generated summaries are from YouTube video captions, while 79.49\% are from Stack Overflow answers. 
We empirically observe that sentences from YouTube videos often discuss the operational steps involved in utilizing the target API, while those operation details are not commonly included in \so answers.
This indicates that the information in YouTube videos can effectively supplement \so in the API documentation augmentation task.

\subsection{Baselines}
\label{sec:baseline}

\subsubsection{Baselines For \compone.\\}
\label{sec:baseline1}
\noindent \textbf{Baseline \#1: SISE} is the state-of-the-art insight sentence extractor~\cite{treude2016augmenting} that identifies insight \so sentences for a given API to augment \doc. 
We modify the classifier in SISE from binary to multi-class.  
\textbf{\\Baseline \#2: DeepTip} is the state-of-the-art approach to identify tip sentences to answer API-related questions from \so answers based on textCNN model~\cite{wang2019extracting}. 
We reuse the architecture of DeepTip with one hot encoding but modify the classifier from a binary classifier to a multi-class classifier.

\subsubsection{Baselines For \extalgo}
\label{sec:baseline2}

\textbf{\\Baseline \#1: TextRank} is a widely used graph-based unsupervised summarization approach~\cite{mihalcea2004textrank}. 
Since our \comptwo is built on top of TextRank, we consider TextRank as our baseline.

\noindent\textbf{Baseline \#2: LexRank}
~\cite{erkan2004lexrank} is an unsupervised graph-based text summarization approach that is similar to TextRank. 
Differently, LexRank leverages inverse document frequency-based cosine similarity to measure the similarity between two sentence nodes. 

\noindent\textbf{Baseline \#2: TechSumBot}
is the SOTA approach for question-answering summarization~\cite{chengran2022answer} in software engineering.
It consists of three sub-modules: 1) the usefulness ranking module to filter out useless sentences to the query, 2) the centrality estimation module to extract summative sentences, and 3) the redundancy removal module to remove redundant sentences. 
We remove the usefulness ranking module of TechSumBot because all the input sentences in this stage are considered useful for API understanding.

\noindent\textbf{Baseline \#3: TechSumBot++}
Since the original TechSumBot only considers redundancies among input sentences, we propose a new variant of TechSumBot by integrating redundancy information between selected sentences and API documentation.
To achieve this goal, we modify TechSumBot's redundancy-reducing component to enable update summarization.
In the redundancy reduction component, TechSumBot iteratively calculates the similarity of each candidate sentence to the one selected in the summary and removes redundant sentences. 
Our modification further calculates the similarity between the candidate sentences and the original API document. 
We then remove redundant candidate sentences from the \doc documentation in each iteration. 
We keep the remaining settings of TechSumBot the same.



\subsubsection{End-To-End Baseline}
\label{sec:end2end}
We set $GPT_{sum}$ (detailed in Section~\ref{sec:howgpt}) as the end-to-end baseline for \toolname.

\vspace{-2mm}
\subsection{Implementation Details}
\subsubsection{\compone}
\label{sec: detailofcompone}

We implement a BERT-based API section classifier and context identifier using \emph{Hugging Face}~\cite{huggingface}, a popular deep-learning library for Transformers. 
We use the `bert-base-uncased' version of the model.
Following the fine-tuning strategy in the BERT paper~\cite{devlin2018bert}, we tune the same key hyper-parameters, i.e., learning rate, batch size, and epoch numbers. 

\vspace{-2mm}
\subsubsection{\comptwo}
We obtain the sentence vector representation using the in-domain sentence representation model proposed by TechSumBot. 
Since there is no clear rule about how many sentences should be considered in the summary of each API section~\cite{treude2016augmenting}, 
we set the abstractive summary length, representing the output sentence length of \toolname, to be equivalent to the length of the ground-truth extractive summaries created in \bench, i.e., they have the same number of sentences.
Moreover, we set the length of the extractive summary (i.e., the output of \extalgo) to be twice that of the abstractive summary, thereby providing \extalgo a certain margin for error.
The information in extractive summaries which is irrelevant to the target API, can subsequently be dropped in the abstractive summary by leveraging GPT's comprehensive understanding of the API documentation learned from in-context learning.
We use the GPT-4 checkpoint (i.e., `gpt-4-0613'). 
For in-context learning, we set sample pairs $N$ as 3 since existing summarization approaches work best on this setting~\cite{zhang2023extractive}. No summaries sampled for in-context learning exists in the test set.

\vspace{-2mm}
\subsection{Automatic Evaluation Metrics}
\label{sec:evaluate}
Following previous work on identifying API-relevant sentences~\cite{treude2016augmenting,wang2019extracting}, we use weighted precision, weighted recall, and weighted F1 for the sentence section classification stage. 

Following existing summarization approaches~\cite{chengran2022answer,kazemi2020biased}, we employ ROUGE as our main automatic evaluation metrics for \extalgo~\cite{lin2004rouge}. 
We report the F1 score for ROUGE-1, ROUGE-2, and ROUGE-L. ROUGE-1 and ROUGE-2 evaluate the unigram and bigram overlap between the generated and ground-truth summaries. 
ROUGE-L evaluates the overlap of the longest common sub-sequence between the generated and ground-truth summaries. 
We use the ROUGE calculation code in the pyrouge library~\cite{pyrouge}.

\section{Evaluation}
This section presents our experiment results as well as the corresponding analysis to answer the following three research questions:

\noindent \textbf{RQ1:} How effective is \compone for sentence section classification?

\noindent \textbf{RQ2:} How effective is \extalgo for update extractive summarization?

\noindent \textbf{RQ3:} How effective is \toolname in augmenting API documentation? 



\subsection{RQ1: Effectiveness in the Sentence Section Classification Stage}
\label{sec:rq1}
\noindent\textbf{Experiment Setting.}
To answer this research question, we compare the performance of \compone and the multi-class classification baselines described in Section~\ref{sec:baseline1} 
for sentence section classification task. 
We use the first-stage result of \bench, i.e., a set of sentences labeled with API sections, to evaluate the performance of all approaches. 
Moreover, we consider the following two evaluation settings: 
\begin{itemize}[leftmargin=*]
    \item \textbf{Cross-API Setting}: 
    We randomly select 75\% of the APIs in \bench, in which the corresponding sentences are used to fine-tune the models. 
    The remaining 25\% of the APIs are used for evaluation.
    
    \item \textbf{Within-API Setting}: We randomly select 75\% of the sentences in \bench for fine-tuning models. The remaining 25\% of the sentences are used for evaluation.
\end{itemize}


\begin{table}[htbp]
\small
        \centering

        \caption{Effectiveness in Section Classification Stage}
        \begin{tabular}{lcccc}
\toprule
\textbf{Setting}                              & \textbf{System} & \textbf{Precision} & \textbf{Recall} & \textbf{F1} \\
\midrule
\multirow{4}{*}{\begin{tabular}[c]{@{}l@{}}Cross-API \\ Setting\end{tabular}}
& SISE & 0.44 & 0.41 & 0.43\\
& DeepTip & 0.52 & 0.51 & 0.51\\
& \compone \small{w/o\;context} & {0.58} & {0.57} & {0.57}\\
& \compone & \textbf{0.63} & \textbf{0.62} & \textbf{0.62}\\
\hline
\multirow{4}{*}{\begin{tabular}[c]{@{}l@{}}Within-API \\ Setting\end{tabular}} 
& SISE & 0.62 & 0.61 & 0.61\\
& DeepTip & 0.66 & 0.64 & 0.65\\
& \compone \small{w/o\;context}& {0.74} & {0.72} & {0.73}\\
& \compone  & \textbf{0.78} & \textbf{0.77} & \textbf{0.77}\\

\bottomrule

        \end{tabular}

        \label{table: firstcomp}
    \end{table}

\noindent\textbf{Result \& Analysis.} 
We report the performance of \compone and selected baselines in Table~\ref{table: firstcomp}.
\emph{\compone \small{w/o\;context}} refers to our sentence section classifier without the context-awareness mechanism.
\textbf{Overall, our \compone outperforms all baselines in both cross-API and within-API settings. 
}Specifically, in the cross-API setting, 
our \compone outperforms the best-performing baseline (i.e., DeepTip) by 21.15\%, 21.57\%, and 21.57\% in terms of precision, recall, and F1 score, respectively.
In the within-API setting, our \compone outperforms the best-performing baseline (i.e., DeepTip) by 18.18\%, 20.31\%, and 18.46\% in terms of precision, recall, and F1 score, respectively. 

By comparing \compone and \compone w/o context, we observe that the context-awareness mechanism improves the performance of \compone in terms of all evaluation metrics.
Specifically, the context-awareness mechanism improves the performance of \compone by 8.62\%, 8.77\%, and 7.02\% in terms of precision, recall, and F1 in the cross-API setting; 5.40\%, 6.94\%, and 5.48\% in the within-API setting.

\subsection{RQ2: Effectiveness in the Update Extractive Summarization Stage}
\noindent\textbf{Experiment Setting.}
To answer this research question, we compare the performance of our extractive update summarization algorithm \extalgo with considered extractive summarization baselines described in Section~\ref{sec:baseline2}. 
We report ROUGE scores to assess the token overlapping between the generated extractive summaries and the ground-truth summaries in \bench.
We consider the following two evaluation settings:

\begin{itemize}[leftmargin=*]
    \item \textbf{Summarization from ground-truth API-relevant sentences}: To assess the theoretical summarization ability of each summarization approach, we consider the output of the first stage in \bench, i.e., ground-truly insightful sentences associated with specific \doc sections, as the input of all summarization approaches. Each approach outputs summaries corresponding to API sections. In this setting, there is no training data since all approaches are unsupervised. 

    \item \textbf{Summarization from \compone output}: To simulate real-world application scenarios, we evaluate the performance of \extalgo by summarizing the output of \compone. 
    Since there is no existing baseline, we combine the best performer in \compone stage and the best baseline of \extalgo as the baseline. We follow the same cross-API setting of RQ1 to split the dataset, train, and test the models. 
\end{itemize}


\begin{table}[htbp]
\small
        \centering
        \caption{ Effectiveness of Summarizing from Ground-Truth API-Relevant Sentences}
        \begin{tabular}{cccc}
        \toprule
        \textbf{System} & \textbf{ROUGE-1}   & \textbf{ROUGE-2}  & \textbf{ROUGE-L}     \\
        \midrule
        LexRank & 0.606 & 0.468 & 0.580\\
        TextRank & 0.640 & 0.528 & 0.620 \\
        TechSumBot & 0.672 & 0.549 & 0.653\\
        TechSumBot++ & 0.694 & 0.579 & 0.674\\
        \extalgo & \textbf{0.795} & \textbf{0.720} & \textbf{0.783}\\
        \bottomrule
        \end{tabular}

        \label{table: summarization}
    \end{table}

\noindent\textbf{Result \& Analysis.} 
In Table~\ref{table: summarization}, we report the result for \extalgo summarizing from ground-truthly API-relevant sentences. 
\textbf{The result shows that \extalgo consistently outperforms all the baselines} in terms of ROUGE-1, ROUGE-2, and ROUGE-L. In particular, compared with the best-performing baseline, i.e., TechSumBot++, our \extalgo achieves better performance by 14.55\%, 24.35\%, and 16.67\% in terms of ROUGE-1, ROUGE-2, and ROUGE-L. Note that the performance of TechSumBot++ is higher than TechSumBot for all the evaluation metrics, which justifies our adjustment of TechSumBot.

Besides, \extalgo outperforms TextRank by 24.22\%, 36.36\%, and 26.29\% in terms of ROUGE-1, ROUGE-2, and ROUGE-L, respectively. 
\extalgo is a variant of the TextRank algorithm. Considering that \extalgo deprioritizes sentences that are semantically similar to the original API documentation, the results demonstrate that \extalgo can effectively reduce the information redundancy between the generated summaries and API documentation. 

\begin{table}[htbp]
\small
    \centering
    \caption{Effectiveness of Summarizing from \compone Output}
        \begin{tabular}{cccc}
        \toprule
    \textbf{System} & \textbf{ROUGE-1}  & \textbf{ROUGE-2} & \textbf{ROUGE-L}    \\
    \midrule
    \begin{tabular}[c]{@{}l@{}}\compone w/o context\\ \& TechSumBot++\end{tabular}  & 0.345 & 0.272 & 0.319\\
    \compone \& \extalgo & \textbf{0.417} & \textbf{0.361} & \textbf{0.398}\\
    \bottomrule
    \end{tabular}
    \label{table: sumcsc}
\end{table}

We also report the performance of \extalgo summarizing from \compone output in Table~\ref{table: sumcsc}.
Specifically, \textbf{\compone \& \extalgo outperforms our construct baseline} by 20.87\%, 32.72\%, and 24.76\% in terms of ROUGE-1, ROUGE-2, and ROUGE-L, respectively. 
The superior performance of \extalgo over the baseline demonstrates the effectiveness of our \extalgo in generating extractive update summaries.



\subsection{RQ3: End-to-End Effectiveness}
\noindent\textbf{Experiment Setting.} 
To answer this research question, we compare the abstractive summarization performance of \toolname with $GPT_{sum}$ detailed in Section~\ref{sec:pilot}.
We conduct a user study to examine the quality of generated summaries.
We follow the user study setting and evaluation metrics used in our empirical study on GPT-4 (detailed in Section~\ref{sec:pilot}).
Specifically, we still ask participants to read all augmented API documentation and rate each of them on a 5-point Likert scale in terms of \textbf{informativeness}, \textbf{relevance}, \textbf{readability}, \textbf{redundancy}, and \textbf{faithfulness}. 
The user study involves four participants, all are PhD students with at least two years of programming experience in Java or Python. 
None of them participated in the benchmark construction process.
We augment APIs in the test set of \bench following the cross-API setting of RQ1.
\begin{table}[htbp]
    \centering
    \caption{End-to-end Effectiveness (5-point Likert scale)}
    \resizebox{\columnwidth}{!}{%
    \begin{tabular}{cccccc}
    \toprule
    \textbf{System} & \textbf{Informativeness.}  & \textbf{Relevance}  & \textbf{Readability} & \textbf{Non-Redundancy} & \textbf{Faithfulness} \\
    \midrule
    $GPT_{sum}$ & 3.6 & 3.3 & 4.4 & 4.5 & 3.6 \\
    \toolname & \textbf{4.1} & \textbf{3.8} & 4.5 & 4.5 & \textbf{4.7} \\
    \bottomrule
    \end{tabular}
    \label{table: end2end}
    }
\end{table}


\noindent\textbf{Result \& Analysis.} 
We show the user study results in Table~\ref{table: end2end}. 
We observe \textbf{\toolname significantly outperforms $GPT_{sum}$ in terms of informativeness, relevance, and faithfulness}  (p < 0.01) by 13.89\%, 15.15\%, and 30.56\%, respectively.
Specifically, \toolname is `very informative in terms of informativeness; between `relevant' to `very relevant' in terms of relevance; and `very faithful' in terms of faithfulness.

We also observe \textbf{\toolname and $GPT_{sum}$ achieves on-par performance in terms of redundancy and readability}. 
The negligible difference in terms of redundancy suggests that our extractive update summarization algorithm, \extalgo, performs on par with GPT-4 itself in identifying and eliminating redundant information.
Note that, we employ update summarization in the Extract Phase rather than the Abstract Phase, which could reduce the risk that GPT-4 potentially alters the semantics of generated summaries and produces hallucinations during update summarization.
Meanwhile, it is reasonable that \toolname performs comparably to $GPT_{sum}$ in terms of readability. This can be attributed to the fact that both approaches leverage the abstractive summarization capabilities of GPT-4.

\section{Threats to Validity}
Threats to internal validity are related to \toolname and the implementation errors of the baselines.
We mitigate this threat by double-checking our code and reusing the replication packages of the baselines.
Meanwhile, we acknowledge that the current version of our dataset may still suffer from the subjective bias of the annotators. 
We plan to expand \bench to mitigate this threat in the future. 
We also acknowledge that \toolname still has threats to presenting factually inaccurate information, which stem from errors in the external resources. 
Nevertheless, the implementation of our data provenance mechanism allows effective verification of factualness by tracing back to the original input.

Threats to external validity are related to the generalizability of our benchmark and experimental results.
Specifically, the \bench dataset is built on Python and Java.
They are two of the most popular programming languages.
In addition, the design of \toolname is programming-language and API agnostic.
\toolname can easily be extended to other programming languages and API types after training with the appropriate data. 

Threats to construct validity are related to evaluation metrics. We use ROUGE, a widely used automatic evaluation metric for text summarization approaches~\cite{chengran2022answer,kazemi2020biased}. 
Furthermore, we use informativeness, relevance, readability, diversity, and faithfulness in our human evaluation,
which are widely used to evaluate SE domain and NLP domain summarization
tasks~\cite{xu2017answerbot, chengran2022answer, huang2020have,bhaskar2023prompted}.


\section{Related Work}

\vspace{0.1cm}\noindent{\bf API Documentation Issues.} 
Researchers often conduct surveys about software document issues with software developers.
Uddin et al.~\cite{uddin2015api} summarized and explored the frequency and severity of ten commonly encountered document issues. 
They highlighted that the most pressing problem is document content issues.
Aghajani et al.~\cite{aghajani2019software} built a taxonomy of software document issues, including 162 issue types. 
In their following work~\cite{aghajani2020software}, they investigated 1) the types of document considered useful for each specific development task and 2) the types of documentation issues considered more relevant to developing practitioners.
Furthermore, researchers also employ mining-based strategies to capture software document issues that developers~\cite{rosen2016mobile} or end users~\cite{khalid2014mobile} frequently discuss. Our work is inspired by previous empirical studies on API documentation issues and aims to augment inadequate API documentation.

\vspace{0.1cm}\noindent{\bf API Document Augmentation.} 
Some approaches enhance the API documentation by integrating usage code examples and corresponding descriptions~\cite{uddin2021automatic,zhang2019enriching}. 
Uddin et al.~\cite{uddin2021automatic} treat pairs of API usage examples and corresponding comments extracted from \so as a new form of API documentation. 
Zhang et al.~\cite{zhang2019enriching} proposed ADECK to extract code examples from \so and map them to a specific API usage scenario. Unlike prior works that focus on providing code examples, our work targets augmenting the natural language part of API documentation, which is considered a complementary approach to the above approaches. 

Many works aim to enrich the textual part of API documentation.
Treude et al.~\cite{treude2016augmenting}  try to identify insightful sentences to specific API from \so to augment Java API documentation. Then DeepTip~\cite{wang2019extracting} extract \so sentences as API tips by training a CNN-based classifier. 
All previous works position the task as a binary classification to classify insightful sentences solely from \so. Differently, our approach positions the task as a multi-class classification to map each sentence to an API document section, followed by an updating summarization task to improve the readability of API documentation. 
In addition, we consider the context of the sentence as well as the structure of \doc. 
Finally, we are the first to leverage the abstraction summarization approaches and GPT in this task. 

\balance

\section{Conclusion and Future Work}
To augment API documentation, this work proposes a two-stage framework \toolname to automatically generate complementary summaries to API documentation based on multiple resources at the section level.
\toolname leverages a BERT-based classifier named \compone to classify sentences from multiple sources into the corresponding API documentation sections. 
Then \compone implements a context-awareness mechanism to adjust the prediction results of context-dependent sentences considering their context.  
Next, \comptwo proposes an extractive update summarization algorithm to generate extractive summaries semantically different from the original API documentation, which is followed by GPT generating abstractive summaries guided by the extractive summaries.
Comprehensive evaluation results on our proposed dataset \bench and human evaluation show that 1) both \compone and \comptwo outperform their baselines by a large margin, and 2) \toolname performs better than the strong baseline GPT. 

In the future, we plan to develop \toolname in the form of a browser plug-in to assist API users for better understanding the target API.
In addition, we plan to explore the potential of other online sources, e.g., API tutorials and technical blogs, to further improve the performance of \toolname.

\vspace{-2mm}
\section{Data Availability}
The replication package of \toolname and \bench is at \small{\url{https://github.com/autumn-city/APIDocBooster}}.


\bibliographystyle{ACM-Reference-Format}
\bibliography{ref}

\end{document}